\documentclass[12pt]{article}
\usepackage{amsfonts}
\usepackage{epsfig,amssymb,euscript}
\usepackage{amsmath,amscd}
\usepackage{slashed}

\numberwithin{equation}{section}
\def\appendix#1{\addtocounter{section}{1}\setcounter{equation}{0}
\renewcommand{\thesection}{\Alph{section}}
\section*{Appendix \thesection\protect\indent \parbox[t]{11.15cm}{#1}}
\addcontentsline{toc}{section}{Appendix \thesection\ \ \ #1}}

\def\bbe{{\bf{e}}}

\def\cG{{\cal{G}}}
\def\bI{{\mathbb{I}}}
\def\bC{{\mathbb{C}}}
\def\bR{{\mathbb{R}}}
\def\sF{{\slashed{F}}}
\def\bE{{\bf{E}}}
\def\bw{{\bar{w}}}

\newcommand{\bea}{\begin{eqnarray}}
\newcommand{\eea}{\end{eqnarray}}

\begin{document}

\begin{titlepage}
\begin{center}

\vspace*{-1.0cm}

\hfill DMUS-MP-19-07
\\
\vspace{2.0cm}

\renewcommand{\thefootnote}{\fnsymbol{footnote}}
{\Large{\bf Real Killing Spinors in Neutral Signature}}
\vskip1cm
\vskip 1.3cm
J. B. Gutowski$^1$  and W. A. Sabra$^2$
\vskip 1cm
{\small{\it
$^1$Department of Mathematics,
University of Surrey \\
Guildford, GU2 7XH, UK.\\}}
\vskip .6cm {\small{\it
$^2$ Physics Department, 
American University of Beirut\\ Lebanon  \\}}
\end{center}
\bigskip
\begin{center}
{\bf Abstract}
\end{center}
 Spinorial geometry methods are used to classify solutions admitting Majorana Killing spinors of the minimal 4-dimensional supergravity in neutral signature, with vanishing cosmological constant and a single Maxwell field strength. Two classes of solutions preserving the minimal amount of supersymmetry are found. The first class admits a null-K\"ahler structure and corresponds to a class of self-dual solutions found by Bryant. The second class admits a null and rotation-free geodesic congruence with respect to which a 
parallel frame can be chosen. Examples of solutions
in the former class are pseudo-hyper-K\"ahler manifolds;
and examples in the latter class include self-dual solutions,
as well as a neutral-signature IWP-type solution. 
\end{titlepage}

\section{Introduction}

Much is known about supersymmetric solutions
of 4-dimensional supergravity. The classification programme
was initiated in work of \cite{Tod:1983pm, Tod:1995jf} for
supergravity theories with Lorentzian signature.
The first classification of solutions in a 5-dimensional theory was constructed in \cite{Gauntlett:2002nw},
by making use of Fierz identity/G-structure analysis.
Further extensions of the 4-dimensional Lorentzian analysis,
using similar methods,
was then done in \cite{Caldarelli:2003pb, Cacciatori:2008ek,
Klemm:2009uw}. Classifications of solutions with minimal supersymmetry in D=11 supergravity were also found 
\cite{Gauntlett:2002fz, Gauntlett:2003wb}.  Other work on the classification of supersymmetric
solutions including more general couplings to hypermultiplets in
4 and 5 dimensions \cite{Meessen:2010fh, Bellorin:2007yp}
and in 6 dimensions using spinorial geometry \cite{Akyol:2012cq, Akyol:2013ana}
and Fierz identity/G-structure methods \cite{Cano:2018wnq}. 
Spinorial geometry techniques have proven to be particularly powerful
for the analysis of supersymmetric solutions. This method exploits the fact that spinors can be written
as differential forms \cite{forms1, forms2}. This is then
applied to classifying supergravity solutions by employing gauge transformations in order to express the Killing spinors in simplified canonical forms, which are then used to solve
the Killing spinor equations. Such techniques were first used
to classify supersymmetric solutions in D=11 supergravity
\cite{Gillard:2004xq}, and have also been applied to heterotic and type II supergravity theories \cite{Gran:2005wu, Gran:2005wn, Gran:2005kg, Gran:2005wf}; see also the review  \cite{Gran:2018ijr}
for a comprehensive description of the applications of
spinorial geometry to the classification programme.

This work was performed for Lorentzian signature supergravity theories. Analogous classifications have also been performed
for Euclidian signature theories in D=4, \cite{euclidean1, euclidean2, euclidean3}, making use of 2-component spinor and spinorial geometry techniques. Einstein-Weyl structures, and
the $SU(\infty)$ Toda equation are among some of the geometric structures associated with such supersymmetric solutions. In contrast, the case of 4D supersymmetric solutions in
neutral signature $(+,+,-,-)$ supergravity theories has received relatively little attention. The analysis of 
parallel spinors in such theories has been performed in
\cite{bryant, md, mdp}, and null-K\"ahler structures
are obtained. A further classification of solutions
in $U(1)$ gauged neutral signature 4D supergravity, with a nonzero cosmological constant, and coupled to a single Maxwell field strength,  was constructed in \cite{Klemm:2015mga}.

In this paper, we  determine, using spinorial geometry techniques, the classification of supersymmetric solutions of the the minimal D=4 neutral signature supergravity, coupled to a Maxwell field strength and
with vanishing cosmological constant. 
The D=4 ungauged neutral signature ${\mathcal{N}}=2$ supergravity theory, coupled to
an arbitrary number of abelian vector multiplets,
with scalars taking values in a projective special para-K\"ahler manifold, was obtained from D=11 M* theory \cite{ch} via a reduction
on $CY_3 \times S^1$ \cite{Sabra:2017xvx}.
In our work, we consider the truncation of this theory
to a single Maxwell field strength, with constant scalars.
The case of minimal $N=1$ supersymmetric solutions is considered, for which the Killing spinors are Majorana. It is shown that there are two 
orbits associated with such spinors, and the simplified canonical forms are obtained using appropriately chosen
$Spin(2,2)$ gauge transformations. For both orbits, the necessary and sufficient conditions for a solution to be supersymmetric are determined. One orbit, for which the spinor is chiral, corresponds to
a sub-class of the solutions considered in \cite{bryant, md, mdp} for which the spinor is parallel with respect to the Levi-Civita connection and the geometry admits a self-dual instanton. However, the geometry associated with the other orbit, for which the spinor is non-chiral,
is novel. Moreover, it does not arise as a limit of
solutions constructed in the analysis of \cite{Klemm:2015mga} for the case of a positive cosmological constant.

The plan of this paper is as follows. In Section 2, the canonical Majorana orbits are determined. In Section 3,
the Killing spinor equations for spinors in the two canonical orbits are analyzed, and the conditions on the geometry and the Maxwell field strength determined. Examples of solutions are also constructed. In Section 4, a ``Wick rotated" Killing spinor equation is considered, and it is shown that solutions
of this Killing spinor equation are in 1-1 correspondence with
those considered in Section 3. In Section 5, we present our conclusions, and discuss the relationship between the non-chiral spinor orbit geometry, and the classification of \cite{Klemm:2015mga}. In Appendix A the spinorial geometry conventions are presented, and in Appendices B and C some further details of the analysis of the Killing spinor
equation conditions in Section 3 are given.

\section{Majorana Spinor Orbits}

The Killing spinor equation (KSE) which we consider is given by
\bea
\label{kse1}
D_\mu \epsilon \equiv \nabla_\mu \epsilon- {1 \over 4}\sF \Gamma_\mu \epsilon =0
\eea
where $F$ is the Maxwell field strength, which satisfies
\bea
dF=0, \qquad d \star F=0 \ .
\eea

In particular, if $\epsilon$ satisfies ({\ref{kse1}}) then so does $C*\epsilon$.
Hence it is sufficient to consider Majorana spinors $\epsilon$ which satisfy $C*\epsilon=\epsilon$. We begin by choosing simple canonical forms for the Majorana spinors.
A basis (over $\bR$) for Majorana spinors $\{\eta_1, \eta_2, \eta_3, \eta_4 \}$, satisfying $C* \eta_i = \eta_i$, is given by
\bea
\label{majsp}
\eta_1 = 1+e_{12}, \quad \eta_2 = i(1-e_{12}), \quad \eta_3 = e_1 +e_2, \quad \eta_4=i(e_1-e_2) \ .
\eea
Consider the spaces of Majorana spinors ${\rm Span}_\bR (\{ \eta_1, \eta_2 \})$
and ${\rm Span}_\bR (\{\eta_3, \eta_4 \})$. We remark that the spinor orbits for the case of $(2,2)$ signature
presented in \cite{bryant} correspond to Majorana spinors, where the spinors are taken to be
in $\bR^2 \oplus \bR^2$. The two copies of $\bR^2$ are identified with  ${\rm Span}_\bR (\{ \eta_1, \eta_2 \})$
and ${\rm Span}_\bR (\{\eta_3, \eta_4 \})$. To evaluate canonical forms for Majorana spinors, we use
only  {\it real} $Spin(2,2)$ gauge transformations which commute with $C*$ and hence preserve the Majorana condition.
Here we use the conventions presented in Appendix A.

The action of $\sigma_1, \sigma_2, i \sigma_3$ associated with
(real) $Spin(2,2)$ gauge transformations, acting independently on 
these vector spaces is generated by
\bea
\label{sl21}
T_1 = \begin{pmatrix} & 1 \quad & 0
\cr & 0 \quad & -1 \end{pmatrix}, \quad T_2 = \begin{pmatrix} & 0 \quad & -1
\cr & -1 \quad & 0 \end{pmatrix}, \quad T_3 = \begin{pmatrix} & 0 \quad & -1
\cr & 1 \quad & 0 \end{pmatrix} \ .
\eea
It follows that a $SL(2,\bR)$ transformation generated by $T_1, T_2, T_3$ can be used to write $\epsilon_1 \in {\rm Span}_\bR (\{ \eta_1, \eta_2 \})$ as $\epsilon_1=a. \eta_1$ where $a=0$ or $a=1$. A similar argument can be used to write $\epsilon_2 \in {\rm Span}_\bR (\{\eta_3, \eta_4\})$ as $\epsilon_2 = b. \eta_3$ for $b=0$ or $b=1$.

So there are three possible canonical Majorana spinors corresponding to
\bea
\epsilon=\eta_1, \quad {\rm or} \quad \epsilon=\eta_3, \quad {\rm or} \quad \epsilon=\eta_1+\eta_3
\eea
however $\eta_1$ and $\eta_3$ are also related by the $Pin(2,2)$ transformation generated by $\gamma_1$, so there
are two canonical Majorana spinors given by
\bea
\label{canon}
\epsilon=1+e_{12}, \quad {\rm or} \quad \epsilon=1+e_{12}+e_1 +e_2 \ .
\eea
We next analyse the conditions obtained from ({\ref{kse1}}) in these two cases.

\section{Analysis of the Killing Spinor Equation}

In this section, we analyse the Killing spinor equation
({\ref{kse1}}). For each of the
two classes of canonical Majorana spinors given in ({\ref{canon}}) we derive the necessary and sufficient conditions on the geometry and the Maxwell field strength.

\subsection{Solutions with Killing Spinor $\epsilon=1+e_{12}$}

Consider first the KSE ({\ref{kse1}}) in the case for which $\epsilon=1+e_{12}$. We find,
on acting with $\gamma_5$ on ({\ref{kse1}}), the conditions
\bea
\nabla_\mu \epsilon=0, \qquad \sF \Gamma_\mu \epsilon=0 \ .
\eea
As the Majorana spinor $1+e_{12}$ is parallel, the geometry  corresponds to one
found in \cite{bryant}. The condition $\sF \Gamma_\mu (1+e_{12})=0$ is also equivalent to $F=\star F$,
where $\epsilon_{1 \bar{1} 2 \bar{2}}=1$.

The bilinear given by

\bea
\chi _{\mu \nu }=i{\mathcal{B}}(\epsilon ,\Gamma _{5}\Gamma _{\mu \nu }\epsilon )
\eea
is represented by the two-form 
\bea
\chi =2\left( \mathbf{e}^{1}\wedge \mathbf{e}^{2}+\mathbf{e}^{\bar{1}}\wedge 
\mathbf{e}^{\bar{2}}\right) +2i\left( \mathbf{e}^{1}\wedge \mathbf{e}^{\bar{1
}}-\mathbf{e}^{2}\wedge \mathbf{e}^{\bar{2}}\right) \ .
\eea
Using the KSE, it can be shown that
\bea
\nabla \chi =0 \ .
\eea
Moreover $\chi $ is null ($\chi ^{2}=0).$ Therefore the solutions admit a null-K\"ahler structure. The metric can be
written in the form \cite{bryant, md, mdp}
\bea
ds^{2}=dwdx+dzdy-S_{x^{2}}dz^{2}-S_{y^{2}}dw^{2}+2S_{xy}dwdz
\eea
where we have used the notation
\bea
S_{x^{2}}=\frac{\partial ^{2}S}{\partial x^{2}},\text{ \ \ \ }S_{y^{2}}=
\frac{\partial ^{2}S}{\partial y^{2}},\text{ \ \ \ }S_{xy}=\frac{\partial
^{2}S}{\partial x\partial y} \ .
\eea
The vanishing of the Ricci curvature implies the conditions:
\bea
\frac{1}{2}S_{x^{2}}S_{xy^{3}}-S_{xy}S_{x^{2}y^{2}}+\frac{1}{2}
S_{y^{2}}S_{x^{3}y}+S_{wx^{2}y}+S_{zxy^{2}}+\frac{1}{2}S_{y^{3}}S_{x^{3}}-
\frac{1}{2}S_{x^{2}y}S_{xy^{2}} &=&0 \nonumber \\
-S_{wxy^{2}}-S_{zy^{3}}-S_{y^{3}}S_{x^{2}y}+(S_{xy^{2}})^{2}-\frac{1}{2}
S_{y^{2}}S_{x^{2}y^{2}}+S_{xy}S_{xy^{3}}-\frac{1}{2}S_{x^{2}}S_{y^{4}} &=&0
\nonumber \\
-S_{wx^{3}}-S_{x^{2}yz}+\left( S_{yx^{2}}\right) ^{2}-S_{xy^{2}}S_{x^{3}}-
\frac{1}{2}S_{y^{2}}S_{x^{4}}+S_{xy}S_{yx^{3}}-\frac{1}{2}
S_{x^{2}}S_{y^{2}x^{2}} &=&0 \ .
\nonumber \\
\eea

\subsubsection{Example: Pseudo-Hyper-K\"ahler metrics}

Further conditions can be obtained if one assumes extended supersymmetry. For example, consider a $N=2$ solution which, 
in addition to the Killing spinor $\epsilon =(1+e_{12})$, also
admits a further Killing spinor given by $\eta =i(1-e_{12}).$

%

The two form spinor bilinear given by

\bea
\left( \chi _{2}\right) _{\mu \nu }=i{\mathcal{B}}(\eta ,\Gamma _{5}\Gamma
_{\mu \nu }\eta )
\eea
is represented by
\bea
\chi _{2}=2i\left( \mathbf{e}^{1}\wedge \mathbf{e}^{\bar{1}}-\mathbf{e}
^{2}\wedge \mathbf{e}^{\bar{2}}\right) -2\left( \mathbf{e}^{1}\wedge \mathbf{
e}^{2}+\mathbf{e}^{\bar{1}}\wedge \mathbf{e}^{\bar{2}}\right) \ .
\eea
Moreover we calculate a third 2-form spinor bilinear given by 
\bea
\left( \chi _{3}\right) _{\mu \nu }=i{\mathcal{B}}(\epsilon ,\Gamma
_{5}\Gamma _{\mu \nu }\eta ) \ .
\eea
This is given by

\bea
\chi _{3}=-2i\left( \mathbf{e}^{1}\wedge \mathbf{e}^{2}-\mathbf{e}^{\bar{1}%
}\wedge \mathbf{e}^{\bar{2}}\right) \ .
\eea
Therefore we have the three bilinears:

\bea
\chi _{1} &=&2\left( \mathbf{e}^{1}\wedge \mathbf{e}^{2}+\mathbf{e}^{\bar{1}
}\wedge \mathbf{e}^{\bar{2}}\right) +2i\left( \mathbf{e}^{1}\wedge \mathbf{e}
^{\bar{1}}-\mathbf{e}^{2}\wedge \mathbf{e}^{\bar{2}}\right) \nonumber \\
\chi _{2} &=&2i\left( \mathbf{e}^{1}\wedge \mathbf{e}^{\bar{1}}-\mathbf{e}
^{2}\wedge \mathbf{e}^{\bar{2}}\right) -2\left( \mathbf{e}^{1}\wedge \mathbf{
e}^{2}+\mathbf{e}^{\bar{1}}\wedge \mathbf{e}^{\bar{2}}\right) 
\nonumber \\
\chi _{3} &=&-2i\left( \mathbf{e}^{1}\wedge \mathbf{e}^{2}-\mathbf{e}^{\bar{1
}}\wedge \mathbf{e}^{\bar{2}}\right) \ .
\eea
We write 
\bea
J_{1} &=&\frac{1}{4}\left( \chi _{1}-\chi _{2}\right) =\left( \mathbf{e}
^{1}\wedge \mathbf{e}^{2}+\mathbf{e}^{\bar{1}}\wedge \mathbf{e}^{\bar{2}
}\right) \nonumber \\
J_{2} &=&\frac{1}{4}\left( \chi _{1}+\chi _{2}\right) =i\left( \mathbf{e}
^{1}\wedge \mathbf{e}^{\bar{1}}-\mathbf{e}^{2}\wedge \mathbf{e}^{\bar{2}
}\right) \nonumber \\
J_{3} &=&\frac{1}{2}\chi _{3}=-i\left( \mathbf{e}^{1}\wedge \mathbf{e}^{2}-
\mathbf{e}^{\bar{1}}\wedge \mathbf{e}^{\bar{2}}\right) \ .
\eea
Then the KSE imply that
\bea
\nabla J_{1}= \nabla J_{2}= \nabla J_{3}=0 
\eea
with
\bea
J_{1}^{2}=-J_{2}^{2}=J_{3}^{2}=1 
\eea
and
\bea
J_1 J_2 = -J_2 J_1 = J_3, \quad J_1 J_3=-J_3 J_1=J_2, \quad J_2 J_3 =-J_3 J_2 = J_1 \ .
\eea
Hence, the geometry admits a pseudo-hyper-K\"ahler metric.

\subsection{Solutions with Killing Spinor $\epsilon=1+e_{12}+e_1+e_2$}

Next, consider the case for which the
Killing spinor is $\epsilon=1+e_{12}+e_1+e_2$. Then the linear system obtained from ({\ref{kse1}}) is as follows:
\bea
\label{linsys}
-\omega_{1,1\bar{1}}+\omega_{1,2\bar{2}}+2i \omega_{1,\bar{1} \bar{2}}&=&\sqrt{2}(-F_{1 \bar{1}}+F_{2 \bar{2}})
\nonumber \\
\omega_{1,1 \bar{1}}-\omega_{1,2 \bar{2}}-2i \omega_{1,12}&=&-2 \sqrt{2}i F_{12}
\nonumber \\
\omega_{1,1 \bar{1}}+\omega_{1,2 \bar{2}}-2i \omega_{1,1 \bar{2}}&=& -2 \sqrt{2}i F_{1 \bar{2}}
\nonumber \\
- \omega_{1,1 \bar{1}}-\omega_{1,2 \bar{2}}+2i \omega_{1, \bar{1} 2} &=&-\sqrt{2}(F_{1 \bar{1}}+F_{2 \bar{2}})
\nonumber \\
-\omega_{2,1\bar{1}}+\omega_{2,2\bar{2}}+2i \omega_{2,\bar{1} \bar{2}}&=&\sqrt{2}i(-F_{1 \bar{1}}+F_{2 \bar{2}})
\nonumber \\
\omega_{2,1 \bar{1}}-\omega_{2,2 \bar{2}}-2i \omega_{2,12}&=&2 \sqrt{2} F_{12}
\nonumber \\
\omega_{2,1 \bar{1}}+\omega_{2,2 \bar{2}}-2i \omega_{2,1 \bar{2}}&=& -\sqrt{2}i (F_{1 \bar{1}}+F_{2 \bar{2}})
\nonumber \\
- \omega_{2,1 \bar{1}}-\omega_{2,2 \bar{2}}+2i \omega_{2, \bar{1} 2} &=&2 \sqrt{2} F_{\bar{1} 2} \ .
\eea

To proceed, consider the 1-form $W$ defined by
\bea
W_\mu = i{\cal{B}}(\epsilon, \Gamma_5 \Gamma_\mu \epsilon)
\eea
and the 2-form $\chi$ given by
\bea
\chi_{\mu \nu} = i {\cal{B}}(\epsilon, \Gamma_5 \Gamma_{\mu \nu} \epsilon) \ .
\eea

These spinor bilinears are given explicitly by
\bea
\label{cc1}
W = 2 \sqrt{2}i (\bbe^1-\bbe^{\bar{1}})-2 \sqrt{2}(\bbe^2+\bbe^{\bar{2}}), 
\qquad
 \chi =W \wedge \theta
\eea
where
\bea
\label{cc2}
\theta = {1 \over \sqrt{2}}(\bbe^1 + \bbe^{\bar{1}}) \ .
\eea
Then
\bea
\label{cc3}
\nabla_\nu W_\mu = {1 \over 2}\eta_{\mu \nu} F_{\lambda_1 \lambda_2}\chi^{\lambda_1 \lambda_2}+ F_{\nu \lambda}\chi^\lambda{}_\mu + F_{\mu \lambda} \chi^\lambda{}_\nu
\eea
and
\bea
\label{cc4}
\nabla_\sigma \chi_{\mu \nu}= F_{\sigma \mu}W_\nu - F_{\sigma \nu} W_\mu
- F_{\mu \nu}W_\sigma +\eta_{\sigma \mu} (i_W F)_\nu - \eta_{\sigma \nu}(i_W F)_\mu \ .
\eea

We remark that the conditions ({\ref{cc1}}), ({\ref{cc2}}), ({\ref{cc3}}), ({\ref{cc4}})
are equivalent to the linear system ({\ref{linsys}).
To proceed with the analysis of these conditions, note that ({\ref{cc3}}) implies that
\bea
\label{cc3b}
dW=0 \ ,  \qquad \nabla^\mu W_\mu=0 \ .
\eea
Furthermore, we also have as a consequence of ({\ref{cc4}})
\bea
\label{cc4b}
W \wedge \big(F+ d \theta \big)=0
\eea
and also
\bea
\label{cc4c}
(i_W F)_\mu = \nabla^\lambda \chi_{\lambda \mu} \ .
\eea

On substituting ({\ref{cc4b}}) and ({\ref{cc4c}}) into ({\ref{cc4}}), it follows that
({\ref{cc4}}) is equivalent to
\bea
\label{cc5}
W_\sigma F_{\mu \nu} = -{1 \over 2} \nabla_\sigma \chi_{\mu \nu}
-{1 \over 2}(W \wedge d \theta)_{\sigma \mu \nu}
+{1 \over 2} \eta_{\sigma \mu} \nabla^\lambda \chi_{\lambda \nu}
-{1 \over 2} \eta_{\sigma \nu} \nabla^\lambda \chi_{\lambda \mu} \ .
\eea
This condition determines all components of $F$ in terms of the geometry.
On using ({\ref{cc5}}) to eliminate $F$ from ({\ref{cc3}}), it follows that
({\ref{cc3}}) is equivalent to 
\bea
\label{part1}
\nabla_W \theta = (\nabla^\nu \theta_\nu) W \ .
\eea

To proceed, consider ({\ref{cc4b}}). This implies that
\bea
F=-d \theta+W \wedge \psi
\eea
for some 1-form $\psi$. The Bianchi identity implies that
\bea
W \wedge d \psi=0
\eea
and hence here exists a function $\cG$ such that
\bea
F=-d \theta + W \wedge d \cG =-d \big(\theta+\cG W \big) \ .
\eea
There is a freedom to make the redefinition
\bea
{\hat{\theta}}=\theta+\cG W
\eea
and we note that  ${\hat{\theta}}^2=1$ and ${\hat{\theta}}$, $W$ are orthogonal. On making this
redefinition, and dropping the $\hat{}$, we take without loss of generality
\bea
F=-d \theta \ .
\eea
In addition, on making use of ({\ref{part1}}), it follows that
\bea
\nabla^\lambda \chi_{\lambda \mu} =-(i_W d \theta)_\mu + (\nabla^\lambda \theta_\lambda) W_\mu
\eea
and so ({\ref{cc4c}}) implies
\bea
\label{dfree}
\nabla^\mu \theta_\mu=0
\eea
hence ({\ref{part1}}) simplifies to
\bea
\label{part1b}
\nabla_W \theta =0 \ .
\eea
It remains to evaluate ({\ref{cc3}}) and also ({\ref{cc5}}). On setting $F=-d \theta$,
({\ref{cc3}}) is equivalent to
\bea
\label{wcond1}
\nabla_\nu W_\mu = \eta_{\mu \nu} \theta^\lambda W^\rho (d \theta)_{\lambda \rho}
+2 (i_W d \theta)_{(\nu} \theta_{\mu)}-2(i_\theta d \theta)_{(\nu}W_{\mu)}
\eea
and ({\ref{cc5}}) is equivalent to 
\bea
\label{tcond1}
{1 \over 2}W_\sigma (d \theta)_{\mu \nu} -W_{[\mu}\nabla_{\nu]} \theta_\sigma
- \bigg(-(i_\theta d \theta)_\sigma W_{[\mu}+\theta_{\sigma} (i_W d \theta)_{[\mu}
-W_\sigma (i_\theta d \theta)_{[\mu} \bigg) \theta_{\nu]}
\nonumber \\
-\eta_{\sigma [\mu} \bigg(\theta_{\nu]} (d \theta)_{\lambda \rho} \theta^\lambda W^\rho
+(i_W d \theta)_{\nu]} \bigg)=0 \ .
\eea
The conditions on $\nabla \theta$ obtained from ({\ref{tcond1}}) can be simplified, making use of
({\ref{dfree}}) and ({\ref{part1b}}) to give
\bea
\label{smptt}
\star (\theta \wedge d \theta) = \nabla_\tau \theta
\eea
where $\tau$ is orthogonal to $\theta$ and $W$, and satisfies $\tau^2=-1$.
Details of this analysis are given in Appendix B.

The condition ({\ref{wcond1}}) is equivalent, together with ({\ref{cc3b}}), to
\bea
\label{ntw}
\nabla_\tau W = \star (W \wedge d\theta)
\eea
and
\bea
\label{framecdt}
\nabla_V W = \star (\tau \wedge d \theta) -i_\theta d \theta
\eea
where, with respect to the frame $\{ V, W, \tau, \theta \}$, the metric is
\bea
\label{newframe}
ds^2=2VW+\theta^2-\tau^2
\eea
with volume form ${\rm dvol}=W \wedge V \wedge \tau \wedge \theta$.
Details of this analysis are given in Appendix C.

Hence, the geometric conditions obtained so far, associated with the frame ({\ref{newframe}}),
can be written as
\bea
d \star \theta=0, \qquad \nabla_W \theta=0, \qquad \nabla_\tau \theta = \star(\theta \wedge d \theta)
\eea
and
\bea
dW=0, \qquad d \star W=0, \qquad \nabla_\tau W = \star (W \wedge d \theta), 
\qquad \nabla_V W = \star (\tau \wedge d \theta)-i_\theta d \theta
\nonumber \\
\eea
and the gauge field is
\bea
\label{Ffield}
F=-d \theta \ .
\eea
We remark that these conditions, as well as the metric ({\ref{newframe}}) are invariant under the redefinitions
\bea
\label{shifty}
V'=V-\beta \tau, \qquad \tau'=\tau-\beta W \ .
\eea
Next we consider the gauge field equations. In order to analyse these,
we note that
\bea
\star \chi = -W \wedge \tau
\eea
and that the condition ({\ref{cc4}}) can be rewritten as
\bea
\nabla_\sigma \star \chi_{\mu \nu} = -W_\sigma \star F_{\mu \nu}
+W_\mu \star F_{\nu \sigma}-W_\nu \star F_{\mu \sigma}
+ \eta_{\sigma \mu} (i_W \star F)_\nu - \eta_{\sigma \nu} (i_W \star F)_\mu
\nonumber \\
\eea
from which it follows that
\bea
d \star \chi = W \wedge \star F
\eea
and therefore
\bea
W \wedge (\star F - d \tau)=0 \ .
\eea
Hence, it follows that there exists a 1-form $\psi$ such that
\bea
\star F = d \tau + W \wedge \psi \ .
\eea
The gauge field equations $d \star F=0$ then imply that
\bea
W \wedge d \psi =0 \ .
\eea
Hence there exists a function ${\cal{H}}$ such that
\bea
\star F = d \tau + W \wedge d {\cal{H}} \ .
\eea
By making use of a redefinition of the form ({\ref{shifty}}) we can
without loss of generality set ${\cal{H}}=0$, and then (dropping the primes)
\bea
\star F = d \tau \ .
\eea
As we have already found the condition $F=-d \theta$, the gauge field equations are
equivalent to
\bea
\label{maxeq}
\star d \theta =- d \tau \ .
\eea
On substituting ({\ref{maxeq}}) into the condition ({\ref{ntw}}), and making use of the
closure of $W$, it follows that ({\ref{ntw}}) is equivalent to
\bea
\nabla_W \tau =0 \ .
\eea
As we also have the conditions $\nabla_W W=0$ and $\nabla_W \theta=0$, and the metric
must also be parallel with  respect to $W$, this also implies that $\nabla_W V=0$
as well, i.e. the frame $\{V, W, \tau, \theta \}$ is parallel with respect to $W$. We also remark that the condition ({\ref{framecdt}}) can be rewritten
as 
\bea
\nabla_V W + \nabla_W V - \nabla_\tau \tau + \nabla_\theta \theta =0 \ .
\eea

It remains to consider the Einstein equations. The integrability conditions of
the KSE, together with the Bianchi identity and gauge field equations, imply that
\bea
\label{einint1}
E_{\mu \nu} \Gamma^\mu \epsilon=0
\eea
where $E_{\mu \nu}=0$ is equivalent to the Einstein equations.
Then ({\ref{einint1}}) implies that
\bea
\label{einint2}
E_{\mu \nu} W^\nu=0 \ .
\eea
Furthermore, ({\ref{einint1}}) implies that
\bea
\bigg(E_{\mu \rho} + \Gamma_\rho{}^\nu E_{\mu \nu} \bigg) \epsilon=0 \ .
\eea
On making use of the condition ${\cal{B}}(\epsilon,\Gamma_5 \epsilon)=0$
and ${\cal{B}}(\epsilon, \epsilon)=0$, the above expression implies that
\bea
\chi_\rho{}^\nu E_{\mu \nu}=0, \qquad {\rm and} \qquad \star \chi_\rho{}^\nu E_{\mu \nu}=0
\eea
or equivalently
\bea
\label{einint3}
E_{\mu \nu} \theta^\nu=0, \qquad E_{\mu \nu} \tau^\nu=0 \ .
\eea
Then ({\ref{einint2}}) and ({\ref{einint3}}) imply that the only component
of the Einstein equations not implied to hold by supersymmetry is
that corresponding to 
\bea
\label{nonein}
V^\mu V^\nu E_{\mu \nu}=0 \ .
\eea

To summarize; in the case for which the
Killing spinor is $\epsilon=1+e_{12}+e_1+e_2$, the necessary and sufficient conditions for supersymmetry are:
\bea
\label{scond1}
d \star \theta=0, \qquad \nabla_\tau \theta = i_\theta (d \tau)
, \qquad  \star d \theta = -d \tau
\eea
and
\bea
\label{scond2}
dW=0, \qquad \nabla_W {\bf{e}}^a=0, 
\qquad \nabla_V W + \nabla_W V - \nabla_\tau \tau + \nabla_\theta \theta  =0
\nonumber \\
\eea
where ${\bf{e}}^a$ denotes the frame $\{V, W, \tau, \theta \}$, with respect to which the metric is
\bea
ds^2=2VW+\theta^2-\tau^2
\eea
and the Maxwell field strength is
\bea
F=-d \theta \ .
\eea
The conditions ({\ref{scond1}}) and ({\ref{scond2}}) imply that $W$ is co-closed, so this condition has been omitted from ({\ref{scond1}}), ({\ref{scond2}}). These conditions are also sufficient to ensure that all components of the Bianchi identity and gauge field equations hold, and that all but one component of the Einstein equations are satisfied. The remaining component of
the Einstein equations which is not implied
by supersymmetry is ({\ref{nonein}}), which must be imposed
as an additional condition to the above.

\subsubsection{Example: Solutions with $F=0$}

For solutions with $F=0$, we note that if $\epsilon$ is
a parallel Killing spinor, then so is $\gamma_5 \epsilon$.
So, taking $\epsilon=1+e_{12}+e_1+e_2$, this implies that
both $1+e_{12}$ and $e_1+e_2$ are independently parallel, i.e. such solutions 
are actually $N=2$ solutions which admit two spinors
$\epsilon_1 = 1+e_{12}$ and $\epsilon_2=e_1+e_2$. These are therefore special cases of the solutions considered in Section 2.1, and have also been considered in \cite{bryant}.

It is however instructive to consider how such solutions appear
in terms of the classification of solutions constructed for
the spinor $\epsilon=1+e_{12}+e_1+e_2$. In particular, the condition $F=0$
together with ({\ref{Ffield}}) and ({\ref{maxeq}}) implies that
\bea
d \theta= d \tau=0
\eea
and hence local co-ordinates $v,u,x,y$ can be introduced such that
\bea
W= du, \quad V=dv+Adu+h_1 dx+h_2 dy, \quad \theta=dx, \quad \tau=dy
\eea
with the vector field dual to $W$ given by $W={\partial \over \partial v}$.
Then 
\bea
{\cal{L}}_W W={\cal{L}}_W \tau={\cal{L}}_W \theta=0
\eea
because $W$, $\tau$ and $\theta$ are closed. Furthermore, the conditions 
\bea
\nabla_V W=\nabla_W V=0
\eea
imply that
\bea
{\cal{L}}_W V=0
\eea
also, so $W$ is an isometry and therefore parallel. In particular, this implies
that the functions $A$, $h_1$, $h_2$ are independent of $v$. A co-ordinate change
of the form $v=v'+f_1(u,x,y)$ can then be used to set, without loss of generality
$h_2=0$, so $V=dv+A du+h dx$ wheren $A$ and $h$ are independent of $v$ (dropping the prime)

Next, the condition $\nabla_\tau \theta=0$ implies that $\partial_y h=0$, so a further co-ordinate
transformation of the form $v=v'+f_2(u,x)$ can be used to set $h=0$ as well. It follows that in these co-ordinates, the metric is
\bea
\label{wavebryant}
ds^2=2du(dv+A du)+dx^2-dy^2 \ .
\eea
It remains to impose the Einstein equations, which are equivalent to $R_{vv}=0$. This implies that
\bea
{\partial^2 A \over \partial y^2}-{\partial^2 A \over \partial x^2}=0 \ .
\eea
Hence, $A$ is a $u$-dependent function which is harmonic on $\bR^{1,1}$, i.e.
\bea
A=A_+(u,x+y)+A_-(u,x-y) \ .
\eea
In particular, the metric ({\ref{wavebryant}}) is self-dual if and only if
\bea
{\partial^2 A_+ \over \partial (x+y)^2}=0
\eea
and the metric is anti-self-dual if and only if
\bea
{\partial^2 A_- \over \partial (x-y)^2}=0 \ .
\eea

Such solutions have already been constructed in \cite{Barrett:1993yn}, albeit in different co-ordinates. In particular, 
the metric is written in terms of co-ordinates $\{p, t, u, v\}$
as
\bea
\label{mmt1}
ds^2=dp dt -{1 \over 2}p^2 du dv -{1 \over 2} p^2 H du^2
\eea
where $H=H(p,u)$. On changing co-ordinates to $\{p, {\hat t}, {\hat u}, {\hat v} \}$,
where 
\bea
{\hat u}=pu, \quad {\hat v}=pv, \quad {\hat t}=t+{1 \over 2}puv
\eea
the metric is
\bea
\label{mmt2}
ds^2 = dp d {\hat t} -{1 \over 2} d {\hat u} d {\hat v}
-{1 \over 2} H (d {\hat u})^2 + p^{-1} {\hat u} H d {\hat u}
dp -{1 \over 2} p^{-2} ({\hat u})^2 H dp^2
\eea
for $H=H(p, {\hat{u}})$. Finally, on setting 
\bea
{\hat t}={\tilde t} + g_1(p, {\hat u}), \qquad
{\hat v}={\tilde v}+ g_2(p, {\hat{u}})
\eea
where $g_1, g_2$ are chosen to satisfy
\bea
{\partial g_1 \over \partial p} = {1 \over 2}p^{-2} ({\hat u})^2 H,
\qquad {1 \over 2}{\partial g_2 \over \partial p} = p^{-1} {\hat u} H + {\partial g_1 \over \partial {\hat u}}
\eea
the metric is
\bea
\label{mmt3}
ds^2 = dp d{\tilde t} -{1 \over 2} d {\hat u} d {\tilde v}
+ {\tilde H} (d {\hat u})^2
\eea
where ${\tilde H}={\tilde H}(p, {\hat u})$ is given by
\bea
{\tilde H}=-{1 \over 2} H -{1 \over 2}{\partial g_2 \over \partial {\hat u}} \ .
\eea
The metric ({\ref{mmt3}}) corresponds to the solution given in 
({\ref{wavebryant}}), in the special case for which the wave profile function $A$ contains only
either left (or right) moving modes, on making a trivial re-labelling of the co-ordinates.

\subsubsection{Example: Solutions with self-dual $F$}

A more general class of solutions is obtained by taking non-zero, but self-dual, $F$. If $F=\star F$, then the geometric conditions
imply that
\bea
\label{sd1}
d (\theta+\tau)=0
\eea
and also
\bea
\label{sd2}
\nabla_{\theta+\tau} \theta =0 \ .
\eea
We can then introduce local co-ordinates $\{v, u, x, y\}$ so that the vector field $W$ is $W={\partial \over \partial v}$, and the 1-forms are
\bea
W&=&du
\nonumber \\
V&=&dv + H du + h_1 dx + h_2 dy
\nonumber \\
\tau+\theta &=& dx
\nonumber \\
\tau-\theta &=& q du +p_1 dx + p_2 dy \ .
\eea
Requiring that $d \star W=0$ implies that $\partial_v p_2 =0$, and hence on defining
\bea
{\hat{y}}=\int p_2 dy
\eea
we can without loss of generality take $p_2=1$.
The conditions $\nabla_W {\bf{e}}^a=0$ further imply that
\bea
\partial_v h_2=0
\eea
and 
\bea
\label{sd3}
\partial_v (h_1+{1 \over 2}q)=0 \ .
\eea
The condition $\partial_v h_2=0$ implies that on making a co-ordinate transformation of the form $v={\hat{v}}+{\cal{F}}(u,x,y)$, we can without loss of generality drop the hat, and set $h_2=0$.

Next, the condition ({\ref{sd2}}) implies that
\bea
\label{sd4}
\partial_y (h_1+{1 \over 2}q)=0 \ .
\eea
The conditions ({\ref{sd3}}) and ({\ref{sd4}}) imply that
\bea
h_1 = -{1 \over 2}q+\psi(u,x) \ .
\eea
By making a co-ordinate transformation of the form
\bea
v={\tilde{v}} + {\cal{G}}(u,x)
\eea
we can without loss of generality set $\psi=0$, and take
$h_1=-{1 \over 2}q$.

After making these simplifications, the metric is
\bea
\label{sdsimpmet}
ds^2=2du(dv+H du -{1 \over 2}qdx)-dx(q du+p_1 dx+dy) \ .
\eea

The remaining geometric conditions are obtained from the condition
$\nabla_a {\bf{e}}^a=0$, which implies
\bea
\label{sd5}
\partial_y q +\partial_v H=0
\eea
and $d \star \theta=0$ implies
\bea
\label{sd6}
\partial_y p_1 -{1 \over 2} \partial_v q=0
\eea
and the condition $d \theta=\star d\theta$ further implies
\bea
\label{sd7}
\partial_x q - \partial_u p_1 + \partial_v \bigg({1 \over 2}q^2+p_1 H \bigg)=0 \ .
\eea
The conditions ({\ref{sd5}}), ({\ref{sd6}}) and ({\ref{sd7}}) have a simple geometric interpretation.
In particular, on defining
\bea
\label{nkstruc}
\omega = du \wedge dx \ ,\qquad  {\rm and} \qquad {\hat{\omega}} = (dv+H du -{1 \over 2}q dx) \wedge (q du + p_1 dx + dy)
\eea
it is straightforward to show that $\omega$ and ${\hat{\omega}}$ define  null almost complex
structures with respect to the metric ({\ref{sdsimpmet}}). 
The conditions ({\ref{sd5}}), ({\ref{sd6}}) and ({\ref{sd7}}) are then equivalent
to
\bea
\nabla \omega=0, \qquad i_W d {\hat{\omega}}=0 \ .
\eea
In particular, the metric admits
a null-K\"ahler structure with respect to $\omega$.
This is to be expected, as the Killing spinor equation
({\ref{kse1}}), together with the self-duality condition
$F=\star F$, imply that the chiral spinor $e_1+e_2$ is parallel with respect to the Levi-Civita connection. 
In contrast, the conditions
({\ref{sd5}}), ({\ref{sd6}}) and ({\ref{sd7}}) are insufficient to imply
integrability of the null almost complex structure ${\hat{\omega}}$.

The conditions ({\ref{sd5}}), ({\ref{sd6}}), and ({\ref{sd7}}) imply
that all components of the Ricci tensor vanish, with the exception of 
$R_{uu}$, where
\bea
R_{uu}&=&2H \partial_v^2 H +4q \partial_y \partial_v H
+4 \partial_y \partial_x H +2 \partial_y \partial_u q +2 \partial_v H \partial_y q
\nonumber \\
&-&2 \partial_v q \partial_y H -4 \partial_y H \partial_y p_1
-2(\partial_y q)^2 -4p_1 \partial_y^2 H \ .
\eea
The condition $R_{uu}=0$ must also be imposed. 

\subsubsection{Example: A Neutral Signature IWP Solution}

A final example is that of a neutral signature IWP-type solution, which has metric
\bea
\label{iwpmet}
ds^2=\lambda^2 \sigma^2 (dt+\phi)^2+{1 \over \lambda^2 \sigma^2}(dx^2-dw d \bw)
\eea
where $t, x$ are real co-ordinates and $w$ is a complex co-ordinate.
$\lambda$ and $\sigma$ are functions of the co-ordinates $x, w, \bw$, and
$\phi=\phi_x dx+\phi_w dw + \phi_\bw d \bw$ is a real 1-form whose 
components are $t$-independent. The 1-form $\phi$ satisfies
\bea
d  \phi &=& {1 \over \lambda^2 \sigma^2} \bigg(-i(\sigma^{-1} \partial_x \sigma
- \lambda^{-1} \partial_x \lambda) dw \wedge d \bw
\nonumber \\
&-&2i(\sigma^{-1} \partial_w \sigma
- \lambda^{-1} \partial_w \lambda) dw \wedge dx
\nonumber \\
&+&2i(\sigma^{-1} \partial_\bw \sigma
- \lambda^{-1} \partial_\bw \lambda) d \bw \wedge dx \bigg)
\eea
and the gauge field is 
\bea
F &=& -{1 \over 2} d \bigg[ (\lambda^2+\sigma^2) (dt + \phi) \bigg]
-{i \over 4} \partial_x \bigg({1 \over \lambda^2}-{1 \over \sigma^2} \bigg)
dw \wedge d \bw 
\nonumber \\
&+&{i \over 2} \partial_w \bigg({1 \over \lambda^2}-{1 \over \sigma^2} \bigg) dx \wedge dw - {i \over 2} \partial_\bw \bigg({1 \over \lambda^2}-{1 \over \sigma^2} \bigg) dx \wedge d \bw
\nonumber \\
\eea
which has dual
\bea
\star F &=& -{1 \over 2} d \bigg[ (\lambda^2-\sigma^2) (dt + \phi) \bigg]
+{i \over 4} \partial_x \bigg({1 \over \lambda^2}+{1 \over \sigma^2} \bigg)
dw \wedge d \bw 
\nonumber \\
&-&{i \over 2} \partial_w \bigg({1 \over \lambda^2}+{1 \over \sigma^2} \bigg) dx \wedge dw + {i \over 2} \partial_\bw \bigg({1 \over \lambda^2}+{1 \over \sigma^2} \bigg) dx \wedge d \bw \ .
\nonumber \\
\eea
The Bianchi identity and gauge field equations imply that $\lambda$, $\sigma$ satisfy
\bea
\label{harm1}
{1 \over 4} \partial_x^2 \bigg({1 \over \lambda^2} \pm {1 \over \sigma^2}\bigg)
-\partial_w \partial_\bw \bigg({1 \over \lambda^2} \pm {1 \over \sigma^2}\bigg)=0 \ .
\eea
In order to show that this solution satisfies the requirements to admit a (non-chiral)
Majorana spinor, we first note that the condition ({\ref{harm1}}) implies
that there exist real functions $\cG_\pm$ of the co-ordinates $x,w,\bw$
which satisfy
\bea
\label{mseq1}
2 \partial_x \cG_\pm -4 \partial_w \cG_\pm = {i \over 2} \partial_w
\bigg({1 \over \lambda^2} \mp {1 \over \sigma^2}\bigg) +{i \over 4} \partial_x
\bigg({1 \over \lambda^2} \mp {1 \over \sigma^2}\bigg) \ .
\eea
The integrability conditions which are sufficient to imply local existence
of such functions are equivalent to ({\ref{harm1}}). We shall use the functions
$\cG_\pm$ to define the basis $W, \theta, \tau, V$ as follows:
\bea
\label{wdef}
W=-4 dx -2 dw -2 d \bw
\eea
\bea
\label{thetdef}
\theta = {1 \over 2} (\lambda^2+\sigma^2)(dt + \phi)+{i \over 4}\bigg({1 \over \lambda^2}-{1 \over \sigma^2}\bigg)\big(dw - d \bw \big)
+ \cG_+(-4 dx -2 dw -2 d \bw)
\nonumber \\
\eea
\bea
\label{taudef}
\tau = -{1 \over 2} (\lambda^2-\sigma^2)(dt + \phi)+{i \over 4}\bigg({1 \over \lambda^2}+{1 \over \sigma^2}\bigg)\big(dw - d \bw \big)
+ \cG_-(-4 dx -2 dw -2 d \bw)
\nonumber \\
\eea
and 
\bea
\label{vdef}
V &=& {1 \over 2} \big(-\cG_+^2+\cG_-^2\big)(-4dx -2 dw -2 d \bw)
-{1 \over 8 \lambda^2 \sigma^2}\big(dx-{1 \over 2}(dw+d \bw) \big)
\nonumber \\
&+& \cG_+ \bigg(-{1 \over 2}(\lambda^2+\sigma^2)(dt+\phi)-{i \over 4}
\bigg({1 \over \lambda^2}-{1 \over \sigma^2} \bigg)\big(dw-d \bw\big) \bigg)
\nonumber \\
&+& \cG_- \bigg(-{1 \over 2}(\lambda^2-\sigma^2)(dt+\phi)+{i \over 4}
\bigg({1 \over \lambda^2}+{1 \over \sigma^2} \bigg)\big(dw-d \bw\big) \bigg) \ .
\eea
The metric given with respect to this basis in ({\ref{newframe}}) 
then corresponds to the neutral IWP metric ({\ref{iwpmet}}).
$W$ is closed, and both $W$ and $\theta$ are co-closed. Furthermore,
the gauge field strength satisfies
\bea
F=-d \theta, \qquad \star F = d \tau \ .
\eea
Furthermore, all the remaining geometric conditions in ({\ref{scond1}}) and ({\ref{scond2}})
hold, as well as the Einstein equations.

We remark that the neutral IWP metric is in fact a $N=2$ Majorana solution.
To see this, it is straightforward to show that the neutral IWP solution
satisfies the Killing spinor equation ({\ref{kse1}}), 
with a Dirac spinor $\eta=\lambda.1+\sigma e_1$. From such a Dirac spinor, one can construct two linearly independent non-chiral Majorana spinors
\bea
\epsilon_1 = \eta + C* \eta, \qquad \epsilon_2 = i (\eta- C* \eta)
\eea
and $\epsilon_1$ (or $\epsilon_2$) is related to the spinor $1+e_1+e_2+e_{12}$
via an appropriately chosen $Spin(2,2)$ gauge transformation.

\section{An Alternative Killing Spinor Equation}

Instead of the Killing spinor equation ({\ref{kse1}}),
one can consider the following alternative Killing spinor equation
\bea
\label{kse2}
D_\mu \epsilon = \nabla_\mu \epsilon- {i \over 4}\sF \Gamma_\mu \epsilon =0 \ .
\eea
In this case, note that if $\epsilon$ satisfies ({\ref{kse2}}) then so does $C* \gamma_5 \epsilon$.
Hence it is sufficient to consider spinors $\epsilon$ which satisfy $C* \gamma_5 \epsilon=\epsilon$. A basis of such spinors over $\bR$ is given by $\{\phi_1, \phi_2, \phi_3, \phi_4 \}$
where 
\bea
\phi_1 = \eta_1, \qquad \phi_2 = \eta_2, \qquad \phi_3 = -i \eta_4, \qquad \phi_4 = i \eta_3 \ .
\eea
The action of $\sigma_1, \sigma_2, i \sigma_3$ associated with (real) Spin(2,2) gauge transformations acting on ${\rm Span}_\bR (\{ \phi_1, \phi_2 \})$ is identical to
the $SL(2,\bR)$ generated by $T_1, T_2, T_3$ in ({\ref{sl21}}).

The independent action of $\sigma_1, \sigma_2, i \sigma_3$ associated with (real) Spin(2,2) gauge transformations acting on ${\rm Span}_\bR (\{ \phi_3, \phi_4 \})$ is given by
\bea
{\hat{T}}_1 = -T_1, \qquad {\hat{T}}_2= -T_2, \qquad {\hat{T}}_3 = T_3 \ .
\eea

So in this case there are three possible canonical spinors corresponding to
\bea
\epsilon=\phi_1, \quad {\rm or} \quad \epsilon=\phi_3, \quad {\rm or} \quad \epsilon=\phi_1+\phi_3
\eea
however $\phi_1$ and $\phi_3$ are also related by the $Pin(2,2)$ transformation generated by $\gamma_3$, followed by rescaling with $i$, so there
are two canonical  spinors given by
\bea
\epsilon=1+e_{12}, \quad {\rm or} \quad \epsilon=1+e_{12}+e_1 -e_2 \ .
\eea

The analysis of the case $\epsilon=1+e_{12}$ produces exactly the same conditions
as for the KSE ({\ref{kse1}}), i.e. it is a geometry of the type found in \cite{bryant}.
It remains to analyse the conditions obtained from ({\ref{kse2}}) in the
case $\epsilon=1+e_{12}+e_1-e_2$. However, in this case, it is straightforward to
prove that the resulting linear system is identical to that given in
({\ref{linsys}}) under the change of holomorphic basis indices
$1 \leftrightarrow 2$ and ${\bar{1}} \leftrightarrow {\bar{2}}$.

Hence the supersymmetric solutions of ({\ref{kse2}}) are in 1-1 correspondence
with the supersymmetric solutions of ({\ref{kse1}}) considered previously. 

\section{Conclusions}

In this paper we have classified the supersymmetric solutions of minimal ungauged 4-dimensional supergravity in neutral signature.
We find two classes of solutions with Majorana Killing spinors:
\begin{itemize}
\item[(i)] The Killing spinor is chiral, and is parallel with respect to the Levi-Civita connection. The gauge field strength is self-dual, and the geometry admits a null-K\"ahler structure.
\item[(ii)] The Killing spinor is not chiral. These geometries admit a rotation-free null geodesic congruence $W$ which is constructed from a vector spinor bilinear, and with respect to which a parallel frame exists. Certain other geometric conditions
given in ({\ref{scond1}}), ({\ref{scond2}}) must also hold.
\end{itemize}
We have also considered a number of examples in these two classes. The geometries in class $(i)$ are a sub-case of
those considered in \cite{bryant}, and their geometric structure
is relatively straightforward to interpret, as all spinor bilinears must
be parallel with respect to the Levi-Civita connection.

In contrast, for the class $(ii)$ solutions, the 2-form bilinear $\chi$ does not define a null almost-complex structure, although
\bea
\omega^\pm = W \wedge (\theta \pm \tau)
\eea
which correspond to the self-dual/anti-self-dual parts of
$\chi$, do define null-almost complex structures. However,
the geometric conditions ({\ref{scond1}}) and ({\ref{scond2}})
do not appear to be sufficient to imply that either of $\omega^+$ or $\omega^-$ is integrable. 

We remark that, for our general classification, we have concentrated solely on solutions preserving the minimal $N=1$
supersymmetry; though for some of the explicit examples we
have considered $N=2$ solutions.  In particular, if $\epsilon$ and $C*\epsilon$ are not proportional over $\bC$, then 
$\eta_1=\epsilon+C*\epsilon$ and $\eta_2=i (\epsilon-C*\epsilon)$
are linearly independent (over $\bR$) Majorana Killing spinors, 
i.e. the solution is really a $N=2$ solution.
Hence, to understand solutions of this theory preserving the minimal $N=1$
supersymmetry, it is necessary and sufficient to consider
the specific case of Majorana Killing spinors. We
leave the classification of the $N=2$ solutions to future work.

Finally, we note that the issue of supersymmetry counting
and Majorana Killing spinors is also of relevance in connection
with the work of \cite{Klemm:2015mga}, in which solutions of
the minimal gauged supergravity in neutral signature, and with nonzero cosmological constant $\Lambda$, were classified.
The classification was undertaken for solutions
with $\Lambda<0$ and $\Lambda>0$ separately. For the case
$\Lambda<0$, the supercovariant derivative does not commute with the charge conjugation operator, and so the theory does
not admit Majorana Killing spinors. However, for $\Lambda>0$,
the supercovariant derivative does commute with charge conjugation, and Majorana Killing spinors should exist.
Furthermore, one might expect (some of) the geometries found in
our work to arise as appropriately tuned limits
of $\Lambda>0$ solutions in from \cite{Klemm:2015mga}.
However, this is not the case{\footnote{In particular, for a non-chiral and Majorana Killing spinor, we have shown that one vector spinor bilinear vanishes, whereas the other one - which corresponds to $W$ - is nonzero. Moreover, all scalar spinor bilinears vanish. By comparison with \cite{Klemm:2015mga}, such
solutions appear to be excluded via an algebraic analysis presented in Appendix E.2 of that work; their 2-form $\Phi$ 
corresponds to the spinor 2-form bilinear $\chi$. However,
we note that there is a mis-identification of the orbits
of the space of 2-forms, with respect to the action of $SO(2,2)$, as given in equation (E.4) of  \cite{Klemm:2015mga}.
Hence, a class of $N=1$ solutions with Majorana non-chiral Killing spinors has inadvertently been excluded from the classification of  \cite{Klemm:2015mga}.}}, and the complete geometric interpretation
of the class $(ii)$ solutions remains elusive. It would be 
of interest to investigate these solutions further.

\setcounter{section}{0}
\setcounter{subsection}{0}

\appendix{Conventions}

We begin with a split signature (pseudo)-holomorphic basis,
i.e. a basis 
\bea
\bbe^1, \quad \bbe^2, \quad \bbe^{\bar{1}}=(\bbe^1)^*,
\quad \bbe^{\bar{2}}=(\bbe^2)^*
\eea
with respect to which the metric is
\bea
\label{cframe}
ds^2 = 2 \bbe^1 \bbe^{\bar{1}}
-2 \bbe^2 \bbe^{\bar{2}} \ .
\eea
With respect to this metric we define
\bea
\Gamma_1=\sqrt{2} i_{e_1}, \quad \Gamma_2 = \sqrt{2} i i_{e_2},
\quad \Gamma_{\bar{1}}= \sqrt{2} e_1 \wedge, 
\quad \Gamma_{\bar{2}}= \sqrt{2} i e_2 \wedge \ .
\eea
The gamma matrices act either on
the space of Dirac spinors,
which consists of the complexified
span of $\{1, e_1, e_2, e_{12}=e_1 \wedge e_2\}$, or the subspace of
Majorana spinors within the space of
Dirac spinors. The canonical orbit elements differ depending on
whether the spinors are Dirac or Majorana.

We will find it useful to also work with a real spacetime basis
$\{ {\hat{\bbe}}^1, {\hat{\bbe}}^2, {\hat{\bbe}}^3, {\hat{\bbe}}^4 \}$
with respect to which the metric is
\bea
\label{reframe}
ds^2 = ({\hat{\bbe}}^1)^2+({\hat{\bbe}}^2)^2-({\hat{\bbe}}^3)^2-({\hat{\bbe}}^4)^2
\eea
and take
\bea
\gamma_1 = {1 \over \sqrt{2}}(\Gamma_1+\Gamma_{\bar{1}}),
\quad \gamma_2= {i \over \sqrt{2}}(\Gamma_1-\Gamma_{\bar{1}})
\eea
and
\bea
\gamma_3= {1 \over \sqrt{2}}(\Gamma_2+\Gamma_{\bar{2}}),
\quad
\gamma_4= {i \over \sqrt{2}}(\Gamma_2-\Gamma_{\bar{2}}) \ .
\eea
With respect to the basis $\{1, e_{12}, e_1, e_2\}$, the gamma matrices
$\gamma_\mu$, $\mu=1,2,3,4$ act as
\bea
\gamma_1 &=& \begin{pmatrix} & 0 \quad & \bI
\cr & \bI \quad & 0 \end{pmatrix},
\quad \gamma_2 = \begin{pmatrix} &0 \quad  \ \ &i \sigma_3
\cr &-i\sigma_3 \quad & 0 \end{pmatrix},
\nonumber \\
\gamma_3 &=& \begin{pmatrix} & 0 \quad & -\sigma_2
\cr & \sigma_2 \quad & 0 \end{pmatrix},
\quad 
\gamma_4 = \begin{pmatrix} & 0 \quad & -\sigma_1
\cr & \sigma_1 \quad & 0 \end{pmatrix}
\eea
with{\footnote{$\sigma_i \sigma_j = \delta_{ij} \bI + i \epsilon_{ijk} \sigma_k$}}
\bea
\sigma_1 = \begin{pmatrix} & 0 \quad & 1
\cr & 1 \quad & 0 \end{pmatrix}, \quad \sigma_2 = \begin{pmatrix} & 0 \quad & -i
\cr & i \quad & 0 \end{pmatrix}, \quad \sigma_3 = \begin{pmatrix} & 1 \quad & 0
\cr & 0 \quad & -1 \end{pmatrix} \ .
\eea

The charge conjugation matrix $C$ is given by
\bea
C =  \begin{pmatrix} & \sigma_1 \quad & 0
\cr & 0 \quad & \sigma_1 \end{pmatrix}
\eea
and satisfies
\bea
[C*, \gamma_\mu]=0
\eea
with
\bea
C* 1=e_{12}, \quad C* e_{12}=1, \quad C*e_1 = e_2, \quad C* e_2 = e_1 \ .
\eea
It is useful to note that
\bea
\gamma_{12} = \begin{pmatrix} & -i \sigma_3  \quad & 0 \cr & 0 \quad & i \sigma_3 \end{pmatrix},
\qquad \gamma_{34} = \begin{pmatrix} & i \sigma_3 \quad & 0 \cr & 0 \quad & i \sigma_3 \end{pmatrix}
\eea
\bea
\gamma_{13} = \begin{pmatrix} & \sigma_2  \quad & 0 \cr & 0 \quad & - \sigma_2 \end{pmatrix},
\qquad \gamma_{24} = \begin{pmatrix} & - \sigma_2 \quad & 0 \cr & 0 \quad & - \sigma_2 \end{pmatrix}
\eea
\bea
\gamma_{14} = \begin{pmatrix} &  \sigma_1  \quad & 0 \cr & 0 \quad & - \sigma_1 \end{pmatrix},
\qquad \gamma_{23} = \begin{pmatrix} & \sigma_1 \quad & 0 \cr & 0 \quad & \sigma_1 \end{pmatrix} \ .
\eea
The action of the (real) $Spin(2,2)$ transformations generated by $\gamma_{12} \pm \gamma_{34}$, $\gamma_{13} \pm  \gamma_{24}$ and $\gamma_{14} \pm \gamma_{23}$ decomposes to (independent) actions of
$\{ \sigma_1, \sigma_2, i \sigma_3 \}$ on the spans of basis elements $\{1, e_{12} \}$ and $\{ e_1, e_2 \}$.

A $Spin(2,2)$ invariant inner product ${\cal{B}}$ is given by
\bea
{\cal{B}} (\epsilon, \eta)= \langle B \epsilon, \eta\rangle
\eea
where
\bea
B = \begin{pmatrix} &  \sigma_3  \quad & 0 \cr & 0 \quad &  \sigma_3 \end{pmatrix}
\eea
satisfies
\bea
B.1=1, \qquad B.e_{12}=-e_{12}, \qquad B.e_1 = e_1, \qquad B.e_2 = -e_2 \ .
\eea
This inner product satisfies
\bea
{\cal{B}}(\epsilon, \gamma_\mu \eta)= {\cal{B}}(\gamma_\mu \epsilon,  \eta)
\eea
and also
\bea
{\cal{B}}(\epsilon, \gamma_{\mu \nu} \eta) = - {\cal{B}}(\gamma_{\mu \nu} \epsilon, \eta) \ .
\eea
Furthermore, if both $\epsilon, \eta$ are Majorana, then
\bea
{\cal{B}}(\epsilon,\eta)=-{\cal{B}}(\eta,\epsilon) \ .
\eea
Also, note that
\bea
\gamma_5 \equiv \gamma_{1234} = -\Gamma_{1 \bar{1} 2 \bar{2}} =  \begin{pmatrix} & \bI  \quad & 0 \cr & 0 \quad & -\bI \end{pmatrix} \ .
\eea
If we set $\epsilon_{1 \bar{1} 2 \bar{2}}=1$ with respect to the complex frame ({\ref{cframe}}), this implies that
$\epsilon_{1234}=-1$ with respect to the real frame ({\ref{reframe}}).

\appendix{Analysis of condition ({\ref{tcond1}})}

In order to analyse the condition ({\ref{tcond1}}), we introduce a local real basis
$\{ \bE^i: i=1,2,3,4 \}$ such that
\bea
\label{nnbasis}
ds^2=2 \bE^1 \bE^2 + 2 \bE^3\bE^4
\eea
with $W=\bE^1$. Then in this basis,
\bea
\theta = \alpha \bE^1 + \bE^3 +{1 \over 2}\bE^4 \ .
\eea
We take the volume form to be ${\rm dvol}=\bE^1 \wedge \bE^2 \wedge \bE^3 \wedge \bE^4$.
In this basis the conditions ({\ref{dfree}}) and ({\ref{part1b}}) are equivalent
to
\bea
\nabla_1 \theta_2 + \nabla_2 \theta_1 + \nabla_3 \theta_4+\nabla_4 \theta_3=0
\eea
and
\bea
\nabla_2 \theta_\mu=0 \ .
\eea
Then ({\ref{tcond1}}) implies that the remaining independent condition
obtained from ({\ref{tcond1}}) is
\bea
\label{aacc1}
\alpha (\nabla_4 \theta_3 - \nabla_3 \theta_4) -{1 \over 2} \nabla_1 \theta_3
+\nabla_3 \theta_1 +\nabla_1 \theta_4 - 2 \nabla_4 \theta_1 =0 \ .
\eea
On making use of this condition, one finds that
\bea
(\theta \wedge d \theta)_{134}=-\nabla_4 \theta_1 +{1 \over 2} \nabla_3 \theta_1
\eea
and hence
\bea
\star (\theta \wedge d \theta)_1 = -(\theta \wedge d \theta)_{134}=\nabla_4 \theta_1 -{1 \over 2} \nabla_3 \theta_1 = (\nabla_\tau \theta)_1
\eea
where $\tau=\bE^3 -{1 \over 2}\bE^4$. The remaining components of ({\ref{smptt}}) follow automatically. Furthermore, we remark that if $\tau$ satisfies $\tau^2=-1$,
and is orthogonal to $W$, $\theta$ with respect to the basis ({\ref{nnbasis}}), then 
the most general expression for $\tau$ is given by
\bea
\tau=\beta \bE^1 \pm \bigg(\bE^3 -{1 \over 2}\bE^4\bigg)
\eea
however the term parallel to $\bE^1$, whose coefficient $\beta$ is not determined uniquely,
gives no contribution to $\nabla_\tau \theta$ appearing in ({\ref{smptt}}), as a consequence of
({\ref{part1b}}).

\appendix{Analysis of condition ({\ref{wcond1}})}

On making use of the conditions on $\nabla \theta$ given in 
({\ref{dfree}}), ({\ref{part1b}}), and ({\ref{smptt}}), the condition
({\ref{wcond1}}) can be rewritten as $dW=0$ together with
\bea
\nabla_W W=0
\eea
and
\bea
\label{cddw}
\nabla_1 W_1 &=& 2 \alpha \nabla_3 \theta_4 + \nabla_1 \theta_3 - 2 \nabla_3 \theta_1
\nonumber \\
\nabla_1 W_3 &=&2 \nabla_3 \theta_4
\nonumber \\
\nabla_1 W_4 &=& \nabla_4 \theta_3
\nonumber \\
\nabla_3 W_3 &=& -2 \nabla_3 \theta_2
\nonumber \\
\nabla_3 W_4 &=&0
\nonumber \\
\nabla_4 W_4 &=& -\nabla_4 \theta_2 \ .
\eea
Note that ({\ref{cddw}}) is equivalent to
\bea
\nabla_\tau W = \star (W \wedge d\theta)
\eea
and
\bea
\label{axlast}
\nabla_1 W_1 = -2 (i_\theta d \theta)_1 \ .
\eea
The condition ({\ref{axlast}}) can be rewritten, on taking
\bea
\tau = \bE^3 -{1 \over 2}\bE^4
\eea
and
\bea
V=\bE^2-\alpha \big(\bE^3+{1 \over 2}\bE^4\big)-{1 \over 2}\alpha^2 \bE^1 \ .
\eea
Then ({\ref{axlast}}) can be rewritten as
\bea
\nabla_V W = \star (\tau \wedge d \theta) -i_\theta d \theta \ .
\eea
The relationship between the frame $\{ \bE^1, \bE^2, \bE^3, \bE^4 \}$ and
$\{ V, W, \theta, \tau \}$ is
\bea
W &=& \bE^1
\nonumber \\
V &=& \bE^2-\alpha \big(\bE^3+{1 \over 2}\bE^4\big)-{1 \over 2}\alpha^2 W
\nonumber \\
\tau &=& \bE^3 -{1 \over 2}\bE^4
\nonumber \\
\theta &=& \alpha \bE^1 + \bE^3 +{1 \over 2}\bE^4
\eea
and hence the metric is
\bea
ds^2=2VW+\theta^2-\tau^2
\eea
with volume form ${\rm dvol}=W \wedge V \wedge \tau \wedge \theta$.

{\flushleft{\textbf{Acknowledgements:}}} JG is supported by the STFC
Consolidated Grant ST/L000490/1. 
The work of WS is supported in part by
the National Science Foundation under grant number PHY-1620505. The authors would like to thank
Maciej Dunajski for useful conversations.

\end{document}